| | |
|---|---|
| *Title:* | Drive Asymmetry and the Origin of Turbulence in an ICF Implosion |
| *Author(s):* | Vincent A. Thomas and Robert J. Kares<br>Applied Computational Physics (XCP) Division<br>Los Alamos National Laboratory |
| *Published in:* | Phys. Rev. Lett. **109**, 075004 (2012)<br>August 17, 2012 |

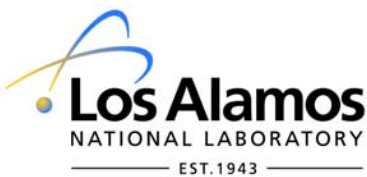

A note to the reader:

Figs. 1(b), 2(a), 2(b) and 2(c) contain embedded hyperlinks to YouTube videos. Please point to these figures and click to play the videos.

# Drive Asymmetry and the Origin of Turbulence in an ICF Implosion

V. A. Thomas and R. J. Kares

*Los Alamos National Laboratory , Los Alamos, New Mexico 87545, USA*

2D and 3D numerical simulations with the adaptive mesh refinement Eulerian radiation-hydrocode RAGE at unprecedented spatial resolution are used to investigate the connection between drive asymmetry and the generation of turbulence in the DT fuel in a simplified inertial-confinement fusion (ICF) implosion. Long-wavelength deviations from spherical symmetry in the pressure drive lead to the generation of coherent vortical structures in the DT gas and it is the three-dimensional instability of these structures that in turn leads to turbulence and mix. The simulations suggest that this mechanism may be an additional important source of mix in ICF implosions. Applications to target ignition at the National Ignition Facility are briefly discussed.

PACS numbers: 52.57.Fg, 47.27.Cn, 47.27.E−, 47.32.C−

Deviations from perfect spherical symmetry in an inertial-confinement fusion (ICF) capsule's pressure drive can arise from many different sources. The impact of these asymmetries on capsule performance has been studied for a number of years because of its potential importance for ICF ignition attempts. Two recent experimental studies are of particular interest.

One set of experiments used spherical plastic (CH) capsules with a $D^3He$ gas fill on the OMEGA laser system and drive asymmetries were induced by offsetting the capsule from the nominal beam focus [1]. In this study the "shock burn" induced by the initial collapse of the converging gas shock on the capsule center was seen to be relatively insensitive to the drive asymmetry while the "compression burn" induced by the compression and heating of the gas by the imploding shell was substantially diminished by increasing drive asymmetry. A second set of experiments with CH capsules on OMEGA showed quantitatively how the amplitude of low mode number ρR asymmetries in the imploded shell at the time of fusion burn are directly correlated with the amplitude of asymmetries in the time-averaged, on-target laser intensity [2]. The authors concluded that the observed growth of low mode ρR asymmetries in the imploded shell are initially seeded by laser illumination asymmetries and grow predominantly by Bell-Plesset (BP) [3] related convergence effects.

In this paper we present results from very high resolution two- and three-dimensional numerical simulations of the implosion of an idealized OMEGA capsule with a simple imposed drive asymmetry. These simulations were performed using the Los Alamos National Laboratory's massively parallel adaptive mesh refinement (AMR) Eulerian radiation-hydrocode RAGE [4] running on 4,096 processors of the US Department of Energy's ASC Purple supercomputer allowing us to achieve unprecedented spatial resolutions as high as 0.05 μm in our 3D implosion simulations which approach but still do not explicitly resolve the Kolmogorov length scale in the system. These simulations were designed to both illuminate the results of the above experiments and to address the more general question of how drive asymmetry leads to turbulence generation in ICF implosions but not to model a specific OMEGA experiment.

The example capsule chosen for our RAGE simulations was a plastic (CH) capsule similar to ones used in the experiments of Rygg et al. [1] on OMEGA and consisted of a round plastic shell with outer radius 425 μm and inner radius 400 μm containing an equimolar mixture of DT gas at an initial density of 2.5 mg/cm$^3$. The drive energy was imparted to the capsule by dividing the CH shell into two regions, an inner region (the pusher) between $r = 400$ μm and 405 μm and an outer region (the ablator) between $r = 405$ μm and 425 μm. Energy was then sourced only into this outer ablator region as an energy source per unit mass $S$ with a fixed spatial profile of the form,

$$S = At[1 + a_\ell P_\ell(\cos\theta)] \tanh[(r - r_0)/\Delta] \quad (1)$$

for $r > r_0$ and $0 < t < 1$ ns where $r_0 = 405$ μm and $\Delta = 20$ μm. Here a value of $A = 1 \times 10^{16}$ ergs/gm/ns was chosen to source in a total of 17.35 kJ in a 1 ns square power pulse. The term proportional to $P_\ell(\cos\theta)$, the Legendre polynomial of order $\ell$, provided a simple asymmetry of amplitude $a_\ell$ for the drive. Although 2D simulations with a variety of values for $\ell$ and $a_\ell$ were performed, the results presented here focus on the case of an energy source with an $\ell = 30$ asymmetry and a relatively large amplitude of $a_{30} = 0.50$ for the asymmetry, i.e. a P30 asymmetry with 50% amplitude. A drive with P30 asymmetry was chosen so that a large number of features are created, which may have some relevance to a laser drive with a large number of beams. The relatively large amplitude for the asymmetry was chosen to more clearly exhibit the mechanism of turbulence growth arising from the imposed drive asymmetry at the relatively low convergence ratio of 8 for the pusher/gas interface achieved in these simulations.

The computational procedure used was to run our capsule simulation from $t = 0$ out to a link time of $t = 1.4$ ns as an axisymmetric 2D RAGE simulation with maximum AMR spatial resolution of 0.4 μm. Then at the chosen link time of $t = 1.4$ ns we rotated one quadrant of the 2D axisymmetric problem into 3D to create a 3D octant version of the axisymmetric data and continued this octant simulation to late time using fully 3D hydro in RAGE.

Fig. 1(a) is an *r-t* plot of the radial motion of the pusher-DT gas interface from a one-dimensional RAGE simulation of our idealized OMEGA capsule which shows the abrupt acceleration, the roughly constant velocity implosion and the stagnation phase with the multiple shock bounces. Also shown is the corresponding radial motion for the main gas shock which suffers its first collapse onto the capsule center at about $t = 1.08$ ns. Stagnation is reached at $t = 1.75$ ns.

Figures 1(b)–1(i) show a series of eight panels, each a time snapshot from the 2D RAGE simulation of the capsule implosion with $a_{30} = 0.50$. Here in the lower portion of each panel the CH shell is colored by pressure while the DT gas is colored by the pressure gradient. In the upper portion of each panel the CH shell is colored by density while the DT gas is colored by the azimuthal component $\omega_\phi$ of the gas vorticity. Figure 1(b) shows the implosion about 0.3 ns after the main shock has broken out into the gas. The asymmetric drive causes non-radial flow in both the pusher and the gas. The non-radial flow in the pusher generates deterministic pressure and density enhancements in the shell at the angular positions of the 15 original drive pressure minima. These perturbations grow in time due to BP [3] related convergence effects [Figs. 1(c)–1(f)] becoming increasingly elongated in the radial direction until shell material protrudes out into the gas. At a later time $t = 1.325$ ns [Fig. 1(g)] the outgoing reflected main gas shock collides with the shell-gas interface and these perturbations are further amplified by the Richtmyer-Meshkov (RM) instability. Subsequent collisions with the outgoing reflected gas shock like the 2nd collision at $t = 1.55$ ns pictured in Fig. 1(i) drive additional RM growth creating large scale deterministic fingers of shell material which penetrate the gas. The result of this process is a well-defined pattern of angular $\rho R$ asymmetries in the pusher shell reminiscent of those observed in the proton spectrometry experiments of Li *et al.* [2].

The non-radial flow in the gas behind the radially inward moving main shock results in the formation of pairs of polar shocks behind the main shock as gas collides in the θ direction at the angular positions of the original drive pressure minima. As the main shock travels radially

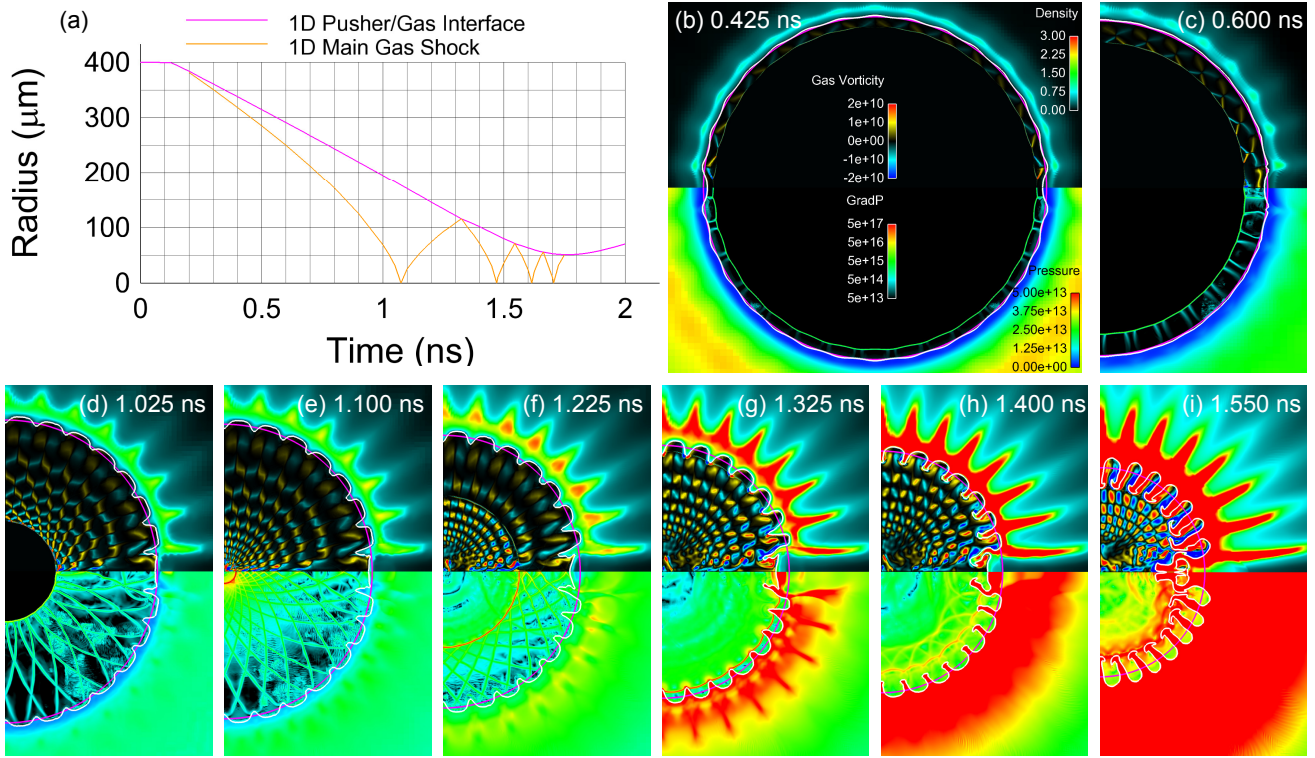

FIG. 1 (color). *r-t* plot (a) from a 1D RAGE simulation of the simplified ICF implosion. Eight time snapshots (b) through (i) taken from a corresponding 2D RAGE simulation with P30 drive asymmetry. The white curve is the CH-gas interface. The magenta circle is the corresponding CH-gas radius in the 1D simulation. In the 2D simulation, the main gas shock remains essentially 1D through the first shock bounce suggesting an explanation for the insensitivity of "shock burn" to drive asymmetry observed by Rygg *et al.* [1].

inward, the moving intersection points between the main shock and these polar shocks formed behind it trace out vortex sheets of opposite sign in $\omega_\phi$ in the gas, a process which can be seen just beginning in Fig. 1(b). This process continues as the main shock converges radially inward toward the center of the capsule and as the polar shocks pass through one another new layers of vortex sheets are created [Figs. 1(c) and 1(d)] so that by the time of the first bounce of the main shock at $t = 1.08$ ns [Fig. 1(e)], several layers of vortex sheets have been formed within the gas. As the outbound main shock passes over these vortex sheets [Fig. 1(f)], it rolls them up into vortex rings of opposite sign. When the outgoing gas shock collides with the shell-gas interface [Fig. 1(g)] at $t = 1.325$ ns, it deposits vortex sheets of opposite sign in $\omega_\phi$ along opposite sides of the developing fingers, a result of the baroclinic torque generated by the oblique collision of the gas shock with the perturbed surface. Subsequent collisions with the outgoing reflected gas shock like the one at $t = 1.55$ ns in Fig. 1(i) deposit even stronger vortex sheets along the sides of the developing RM fingers. As the fingers grow, this vorticity is rolled up into vortex rings of opposite sign in $\omega_\phi$ which are trapped inside the developing bubbles of gas between the growing RM fingers. At link time t = 1.4 ns [Fig. 1(h)], counter-rotating vortex rings are observed in close proximity both in the body of the gas and in the bubbles of gas between the growing RM fingers. Such vortex rings are subject to the unstable growth of azimuthal waves in 3D [5–7].

The 3D RAGE simulation was initialized at $t = 1.4$ ns with an AMR resolution in 3D of 0.20 μm in the region of CH-gas interface. At $t = 1.5$ ns this AMR resolution was increased to 0.10 μm. At $t = 1.6$ ns the AMR resolution was further increased to a final value of 0.05 μm. In this AMR strategy the finest spatial resolution was only utilized at late time when the gas volume was smallest but the resulting total cell count at $t = 1.71$ ns still reaches nearly $1 \times 10^9$ AMR cells. Utilizing this same spatial resolution at earlier times was not practical due to the much larger gas volume to be zoned and as a result 3D turbulent development which might have begun earlier in time could be artificially suppressed in this simulation.

Figures 2(a)–2(d) are a sequence of four time snapshots from the 3D RAGE simulation that show a close up view of the late time evolution of the vorticity in the gas bubbles nearest the polar axis of the capsule illustrating how the growth of azimuthal instabilities on the strong, counter-rotating vortex rings present in the gas bubbles leads to turbulence in the bubbles and the resulting interpenetration of CH and gas. In each of these snapshots only the central gas region of the 3D octant is shown with the surrounding CH shell material removed. The white surface represents the CH-gas interface. The vertical face of the gas is colored by $\omega_\phi$ and the horizontal face is colored by the gradient of pressure. The total vorticity in the interior of the gas is displayed in volume-rendered representation with a transfer function for color and opacity chosen to visualize regions of the flow with total vorticity

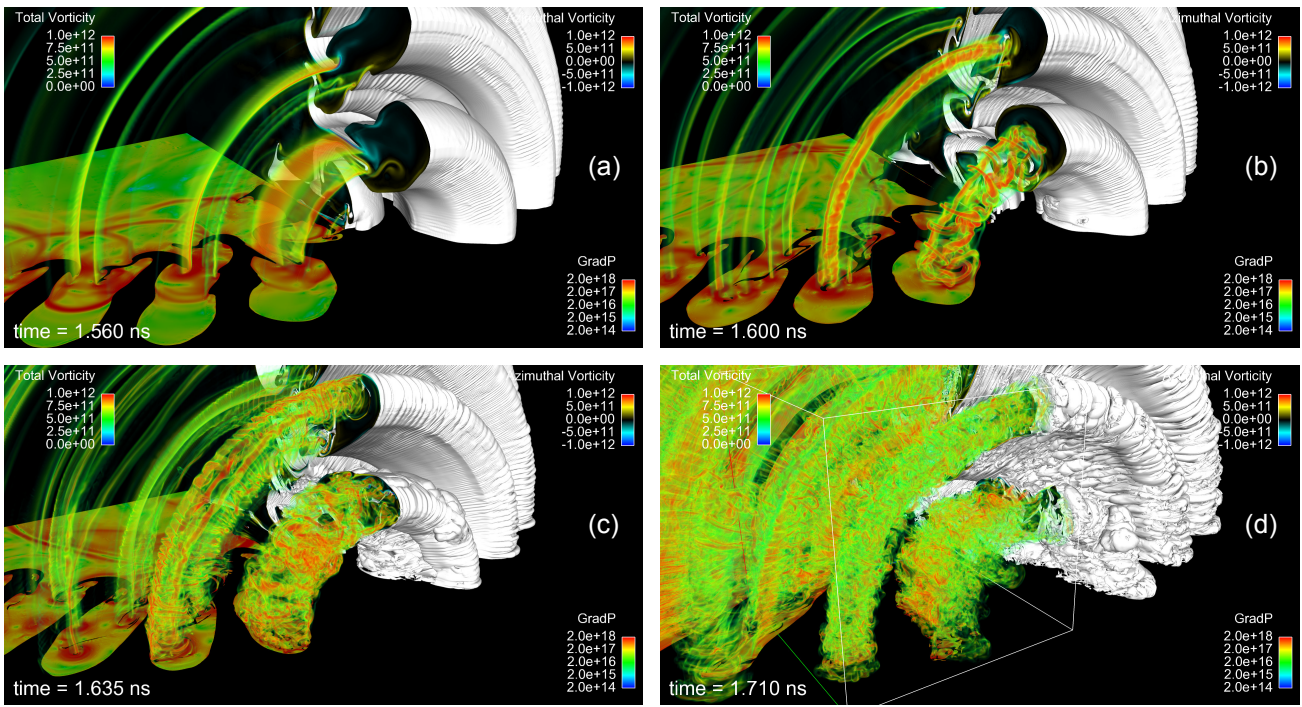

FIG. 2 (color). Four time snapshots from the 0.05 μm 3D RAGE simulation of the late time implosion of the idealized ICF capsule. (a) through (d) show a close up view of the rapid development of the turbulence in the gas bubbles nearest the polar axis of the capsule.

above $5\times10^{11}\,\text{sec}^{-1}$, a technique which is effective in revealing the dynamical evolution of the vortex cores in the interior of the gas bubbles. In Fig. 2(a) vortex sheets of opposite sign can be seen along opposite sides of the gas bubble nearest the polar axis. These sheets are pushed together by the radial convergence of RM fingers and are quickly rolled up into counter-rotating vortex rings that are trapped in the bubble and that immediately begin to undergo azimuthal instability growth in 3D [Fig. 2(b)] with a long wavelength Crow instability [6] clearly evident. This instability first appears in the interior of the gas and it is only when the increasingly turbulent vorticity of the gas flow impacts the CH surface [Figs. 2(c) and 2(d)] that fully 3D interpenetration and mixing occurs. This process may be envisioned as the injection of a gas jet into the bubble though the constriction caused by convergence of the RM fingers and the impact of this turbulent jet on the CH shell wall. A similar process can be seen occurring in the other bubbles as well with azimuthal instabilities of the short wavelength Widnall [5], long wavelength Crow [6] and combined type [7] all contributing to the very rapid evolution of the coherent vortical structures trapped in the bubbles to a fully turbulent state.

Figure 2(d) shows the final timestep at $t = 1.71$ ns with a superimposed square sample box 33 μm on a side that completely encloses the two turbulent bubbles nearest the polar axis. From this sample region we extracted the power spectrum of the total kinetic energy for the 0.05 μm simulation plotted in Fig. 3 together with spectra from two corresponding 3D simulations with lower resolutions of 0.20 μm and 0.10 μm respectively. For resolutions of 0.10 μm and above clear evidence of an inertial subrange is seen in these spectra with Kolmogorov $k^{-5/3}$ scaling. The evidence of fully developed turbulence prior to the time of stagnation is significant for ICF applications because it demonstrates how asymmetry can lead to hydrodynamic turbulence by way of instability of large coherent features in a time that is short enough to be of interest for degrading the compression burn of an ICF implosion.

Our RAGE simulations serve to clarify the essential role of convergence and shear in an asymmetrically driven ICF implosion. Pressure drive asymmetries on the capsule lead to non-radial flow in the shell which in turn leads to macroscopic density enhancements in the shell that grow in time as a result of BP related convergence effects. In the compressible case considered here the BP growth of the density perturbations leads to both radially inward and outward extension of the shell. These macroscopic perturbations act as seeds for the development of RM fingers of shell material in the gas as a result of multiple interactions with the reflected gas shock. This same interaction results in strong shear flow near the fingers in the form of counter-rotating vortical structures in the gas. Radial convergence pushes both the fingers and their associated vortical flow closer together as the implosion progresses. This radial convergence guarantees that the counter-rotating vortical structures will suffer azimuthal instability growth in three dimensions resulting in rapid development of turbulence in the bubbles trapped between the fingers. It is this turbulent growth which leads to interpenetration of shell material and gas that represents fully three-dimensional mixing. The physical picture described above suggests why it is particularly difficult to achieve the ideal 1D yield in a high convergence ICF implosion. Although the simulation shown here used a 2D asymmetry, similar processes would be expected for the case of fully 3D asymmetries.

It has been commonly supposed that yield degradation in the National Ignition Facility ignition target is dominated by large $\ell$ mode Rayleigh-Taylor instabilities at the interfaces [8,9]. Our RAGE simulations suggest, however, an additional possibility. For very high convergence targets like the NIF ignition target with a convergence ratio of 30, even small asymmetries in the capsule drive can lead to penetration of the central DT hotspot by macroscopic fingers of cold fuel from the DT ice layer. The same drive asymmetry that leads to these fingers also leads to the coherent vortical structures in the gas whose 3D instability generates turbulent mixing of the hotspot gas with the cold fuel which may hydrodynamically disrupt the hotspot before stagnation and result in ignition failure.

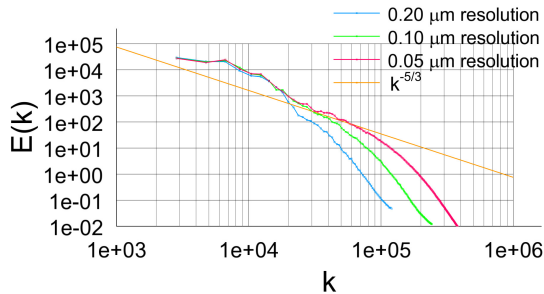

FIG. 3 (color). The power spectra of the kinetic energy at $t = 1.71$ ns extracted from the sample box of Fig. 2(d) for three different simulation resolutions demonstrate the emergence of an inertial subrange with Kolmogorov scaling.